%% file: emask-final.backup.tex
\documentclass[12pt]{article}
\usepackage{times,boxedminipage,epsfig,psfig,alltt,subfigure}

\newcommand{\comment}[1]{}
\newcommand{\COMMENT}[1]{}
\newcommand{\jhcomment}[1]{}
\newcommand{\shcomment}[1]{}

\def\choose#1#2{ \ ^{#1}C_{#2} }

\begin{document}
\title{On Addressing Efficiency Concerns in Privacy-Preserving Mining 
}

\author{
\begin{tabular}{c c c}
Shipra Agrawal$^\dagger$ & 
Vijay Krishnan$^\ddagger$ & 
Jayant R. Haritsa$^\dagger$
\end{tabular}
\\
\small
\begin{tabular}{c c c}
$^\dagger$ Computer Science \& Automation 
& \hspace*{0.2in} & 
$^\ddagger$Computer Science \& Engineering 
\\
Indian Institute of Science & & Indian Institute of Technology \\
Bangalore 560012, INDIA & & Mumbai 400076, INDIA 
\end{tabular}
}


\input{cover}

\maketitle

\setcounter{page}{1}

\begin{abstract}

Data mining services require accurate input data for their results to be
meaningful, but privacy concerns may influence users to provide spurious
information.  To encourage users to provide correct inputs, we recently
proposed a data distortion scheme for association rule mining that
simultaneously provides both privacy to the user and accuracy in the
mining results.  However, mining the distorted database can be 
orders of magnitude more time-consuming as compared to mining the
original database.  In this paper, we address this issue and demonstrate
that by (a) generalizing the distortion process to perform
symbol-specific distortion, (b) appropriately chooosing the distortion
parameters, and (c) applying a variety of optimizations in the
reconstruction process, runtime efficiencies that are well within an
order of magnitude of undistorted mining can be achieved.  

\comment{
We also
discuss why it appears unlikely that efficient and accurate mining
algorithms can be designed for the alternative notions of privacy
proposed in the recent literature.
}

\jhcomment{Check out last sentence}

\end{abstract}

\section{Introduction}
The knowledge models produced through data mining techniques are
only as good as the accuracy of their input data.  One source of
data inaccuracy is when users deliberately provide wrong information.
This is especially common with regard to customers who are asked to
provide personal information on Web forms to e-commerce service providers.
The compulsion for doing so may be the (perhaps well-founded) worry that
the requested information may be misused by the service provider to harass
the customer.  As a case in point, consider a pharmaceutical company that
asks clients to disclose the diseases they have suffered from in order
to investigate the correlations in their occurrences -- for example,
``Adult females with malarial infections are also prone to contract
tuberculosis''. While the company may be acquiring the data solely for
genuine data mining purposes that would eventually reflect itself in
better service to the client, at the same time the client might worry
that if her medical records are either inadvertently or deliberately
disclosed, it may adversely affect her employment opportunities.

Recently, in \cite{mask}, we investigated (for the first time)
whether customers can be
encouraged to provide correct information by ensuring that the mining
process cannot, with any reasonable degree of certainty, violate their
privacy, but at the same time produce sufficiently accurate mining results.
The difficulty in achieving these goals is that privacy and accuracy are
typically contradictory in nature, with the consequence that improving
one usually incurs a cost in the other~\cite{agra01}.

Our study was carried out in the context of extracting \emph{association
rules} from large historical databases, a popular mining
process~\cite{ais93} that identifies interesting correlations between
database attributes, such as the one described in the pharmaceutical
example. For this framework, we presented a scheme called {\bf MASK},
(Mining Associations with Secrecy Konstraints), based on a
simple probabilistic distortion of user data, employing random numbers
generated from a pre-defined distribution function. It is this distorted
information that is eventually supplied to the data miner, along with a
description of the distortion procedure.  A special feature of MASK is
that the distortion process can be implemented at the data source itself,
that is, at the \emph{user machine}. This increases the confidence of
the user in providing accurate information since she does not have to
trust a third-party to distort the data before it
is acquired by the service provider.

Experimental evaluation of MASK on a variety of synthetic and real
datasets showed that, by appropriate setting of the distortion parameters,
it was possible to simultaneously achieve a high degree of privacy and
retain a high degree of accuracy in the mining results.  While these
results were very encouraging, a problem left unaddressed was
characterizing the \emph{runtime efficiency} of mining the distorted data
as compared to directly mining the original data. That is, the question
``Is privacy-preserving mining of association rules more expensive than 
direct mining?" was not considered in detail.
Our subsequent analysis has shown that this issue is indeed a
\emph{major concern}: Specifically, we have found that on typical
market-basket databases, privacy-preserving
mining can take as much as \emph{two to three orders of
magnitude} more time as compared to direct mining.  Such
enormous overheads raise serious questions about the viability of
supporting privacy concerns in data mining environments.

In this paper, we address the runtime efficiency issue in
privacy-preserving association rule mining, which to the best of our
knowledge has never been previously considered in the literature.
Within this framework, our contributions are the following:

\begin{itemize}
\item 
We show that there are inherent reasons as to why mining the distorted
database is a significantly harder exercise as compared to directly
mining the original database.

\item
We demonstrate that it is possible to bring the efficiency to \emph{well
within an order of magnitude} with respect to direct mining, while
retaining satisfactory privacy and accuracy levels.  This improvement is
achieved through changes in both the distortion process and the mining
process of MASK, resulting in a new algorithm that we refer to as EMASK.
In EMASK, the distortion process is generalized to perform
\emph{symbol-specific} distortion -- that is, different distortion
parameters are used for 1's and 0's in a transaction.(1 indicates the presence of an item 
and 0 indicates the absence of an item in a transaction). 
Estimation procedures are designed to carefully chose the parameters of distortion beforehand
and a variety of optimizations are applied in the mining process to achieve the desired goals.  
Our new design is validated against a variety of synthetic and real datasets.

\item

In MASK,  privacy is measured with respect to the ability to directly
reconstruct entries in the distorted matrix. We refer to this hereafter
as ``basic privacy''.  In EMASK, we consider basic privacy as well as
``reinterrogated privacy'' -- this latter notion captures the situation
wherein the miner is permitted to utilize the \emph{mining output}
(i.e. the association rules) to reinterrogate the distorted matrix,
potentially resulting in reduced privacy.

\comment{
Both MASK and EMASK support a notion of ``average privacy'', that is,
they compute the probability of being able to accurately reconstruct a
\emph{random} entry in the database.  An alternative characterization of
privacy proposed recently in \cite{gehrke02} is to evaluate the probability
of accurately reconstructing a \emph{specific} entry in the database.
We discuss here why it appears unlikely that efficient mining algorithms
can be designed to support this stronger measure of privacy.
}

\jhcomment{Check out above para}


\end{itemize}

\comment{
This paper extends the same concern to Association Rule mining. We
assume that the customers are willing to provide distorted values for
their entries. We define a possible mechanism by which the end users
can distort their values and then quantify the amount of privacy they
get in terms of the parameters used in distortion. 
}

\subsection{Organization}
The remainder of this paper is organized as follows: In
Section~\ref{sec:frame}, we describe the privacy framework employed in
our study, while Section~\ref{sec:mask} provides background information 
about the original MASK
algorithm.  Then, in Section~\ref{sec:emask}, we present the details
of our new EMASK privacy-preserving scheme.  The performance model and
the experimental results are highlighted in Sections~\ref{sec:perf} and
\ref{sec:expt}, respectively.  
Related work on privacy-preserving mining is reviewed
in Section~\ref{sec:related}.  Finally, in Section~\ref{sec:conc},
we summarize the conclusions of our study.

\section{Problem Framework}
\label{sec:frame}

\subsection{Database Model}
We assume that each customer contributes a tuple to the database with the
tuple being a fixed-length sequence of 1's and 0's.  A typical example of
such a database is the so-called ``market-basket'' database~\cite{ais93}
wherein the columns represent the items sold by a supermarket, and
each row describes, through a sequence of 1's and 0's, the purchases
made by a particular customer (1 indicates a purchase and 0 indicates
no purchase).  We also assume that the overall number of 1's in the
database is significantly smaller than the number of 0's -- this is
especially true for market-baskets since each customer typically buys
only a small fraction of all the items available in the store.  In short,
the database is modeled as a \emph{large disk-resident two-dimensional
sparse boolean matrix}.

Note that the boolean representation is only logical and that the
database tuples may actually be physically stored as ``item-lists'',
that is, as an ordered list of the identifiers of the items purchased
in the transaction.  The list representation may appear preferable
for the sparse databases that we are considering, since it reduces the
space requirement as compared to storing entire bit-vectors.  However,
because we are \emph{distorting} user information, it
may be the case that the distorted matrix will not be as sparse as the
true database.  Therefore, in this paper, we assume that the distorted
database is stored as a large collection of bit-vectors.

\subsection{Mining Objectives}
The goal of the miner is to compute \emph{association rules} on the above
database.  Denoting the set of transactions in the database by ${\cal T}$
and the set of items in the database by $\cal I$, an association rule
is a (statistical) implication of the form $ \cal X \Longrightarrow Y$,
where $ {\cal X, \cal Y \subset \cal I} $ and ${\cal X} \cap Y = \phi$.
A rule $ \cal X  \Longrightarrow Y $ is said to have a \emph{support}
(or frequency) factor \emph{s} iff at least ${s}\% $ of the transactions
in ${\cal T}$ satisfy $\cal X \cup Y$.  A rule $\cal X  \Longrightarrow
Y$ is satisfied in the set of transactions $ {\cal T} $ with a {\em
confidence} factor \emph{c} iff at least $ c\% $ of the transactions in $
{\cal T} $ that satisfy $ \cal X $ also satisfy $ \cal Y $.  Both support
and confidence are fractions in the interval [0,1].  The support is a
measure of statistical significance, whereas confidence is a measure of
the strength of the rule.

A rule is said to be ``interesting'' if its support and confidence are
greater than user-defined thresholds ${sup}_{min}$ and ${con}_{min}$,
respectively, and the objective of the mining process is to find all such
interesting rules.  It has been shown in \cite{ais93} that achieving this
goal is effectively equivalent to generating all subsets $\cal X$ of $
\cal I $ that have support greater than ${sup}_{min}$ -- these subsets
are called \emph{frequent} itemsets.  Therefore, the mining objective
is, in essence, to efficiently discover all frequent itemsets that are
present in the database.

\subsection{Privacy Metric}
As mentioned earlier, the mechanism adopted in this paper for achieving
privacy is to \emph{distort} the user data before it is subject to the
mining process.  Accordingly, we measure privacy with regard to the
probability with which distorted entries can be \emph{reconstructed}.
In short, our privacy metric is: ``With what probability can a random
1 or 0 in the true matrix be reconstructed''?  
For the sake of presentation simplicity, we will assume in the
sequel that the user is only interested in privacy for 1's (this
appears reasonable to expect in a market-basket environment, since
the 1's denote specific actions, whereas the 0's are default options).
The generalization to providing privacy for both 1's and 0's is
straightforward -- the details are given in \cite{tech-report}.

Privacy can be computed at two levels: \emph{Basic Privacy (BP)} and
\emph{Re-interrogated Privacy (RP)}. In basic privacy, we assume that
the miner does not have access to the distorted data matrix after the
completion of the mining process.  Whereas, in re-interrogated privacy,
the miner can use the mining output (i.e. the association rules) to
subsequently \emph{re-interrogate} the distorted database, possibly
resulting in reduced privacy.

\section{Background on MASK}
\label{sec:mask}

We present here background information on the MASK algorithm, which
we recently proposed in \cite{mask} for providing acceptable
levels of both privacy and accuracy.

Given a customer tuple with 1's and 0's, the MASK algorithm has a simple
distortion process:   Each item value (i.e. 1 or 0) is either kept
the same with probability $p$ or is flipped with probability $1-p$.
All the customer tuples are distorted in this fashion and make up
the database supplied to the miner -- in effect, the miner receives a
\emph{probabilistic function} of the true customer database.  For this
distortion process, the Basic Privacy for 1's was computed to be
\begin{equation}
\textstyle
{{\cal R}_{1}(p)  =  
\frac{s_{0} \times p^{2}}{s_{0} \times p + (1-s_{0}) \times (1-p)} +
\frac{s_{0} \times (1-p)^{2}}{s_{0} \times (1-p) + (1-s_{0}) \times p}}
\label{eq:recon}
\end{equation}
\noindent
where $s_0$ is the average support of individual items in the database and 
$p$ is the distortion parameter mentioned above.

Since the privacy graph as a function of $p$ has a large range where
it is almost constant, it means that we have considerable
flexibility in choosing the $p$ value -- in particular, we can choose
it in a manner that will \emph{minimize the error} in the subsequent
mining process. Specifically, the experiments in \cite{mask} showed
that choosing $p=0.9$ (or, equivalently, $p=0.1$, since the graph is
symmetric about $p=0.5$) was most conducive to accuracy.

The concept of Re-interrogated Privacy was not considered in
\cite{mask}; we include it in this paper and compute both Basic Privacy
and Re-interrogated Privacy for EMASK (see Section~\ref{sec:priv} for
details).


\comment{
A related issue here is whether the user would want the \emph{same}
level of privacy for both 1's and 0's? For many applications, such as the
market-basket database, it appears reasonable to expect that customers
would want more privacy for their 1's than for their 0's, since the 1's
denote specific actions whereas the 0's are the default options.
}

\section{The EMASK Algorithm}
\label{sec:emask}

\begin{figure}[!ht]
\begin{center}
\includegraphics[width=0.4\textwidth]{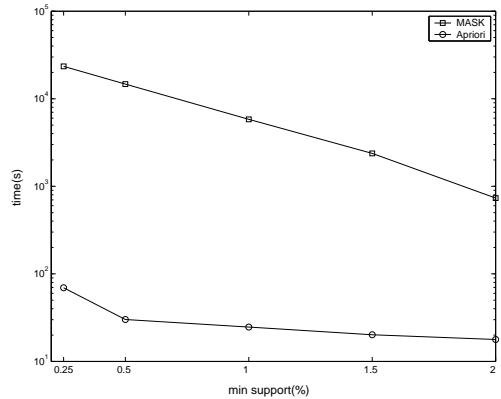}
\end{center}
\caption{Comparison of run time of Apriori and MASK  (log scale)
\label{fig:efficiency}}
\end{figure}

Our motivation for the EMASK algorithm stems from the fact that while
MASK is successful in obtaining the dual objectives of privacy and
accuracy, its runtime efficiency proves to be rather poor. 
To motivate the cause for this inefficiency,
let us look at the performance numbers of a sample database.
Figure \ref{fig:efficiency} shows the running time (on log scale) of MASK 
as compared to Apriori for a synthetic database generated with IBM Synthetic Data 
generator (T10.I4.D1M.N1K as per naming convention of \cite{as94}).
The figure shows that there are huge differences in running times of
the two algorithms --  specifically, mining the distorted database can take as much
as \emph{two to three orders of magnitude} more time than mining the
original database.  Obviously, such enormous overheads make it difficult
for MASK to be viable in practice.

\noindent
The reason for the huge overheads are the following:  
\begin{description}
\item[Increased database density:]~
This overhead is inherent to the methods employing random distortion
method to achieve privacy. The random perturbation methods flip 0's
to 1's to hide the original 1's. Due to the generation of false 1's,
the \emph{density} of the database matrix is increased considerably. For
example, given a supermarket database with an average transaction length
of 10 over a 1000 item inventory, a flipping probability as low as 0.1
increases the average transaction length by an order of magnitude,
i.e. from 10 to 108. The reason that flipping of true 1's to false
0's cannot compensate for this increase is that the datasets we are
considering here are sparse databases, with the number of 0's orders of
magnitude larger than the number of 1's. Hence the effect of flipping
0's highly dominates the effect of flipping 1's.

As a result of increased transaction lengths, counting the itemsets
in the distorted database requires much more processing as compared
to the original database, substantially increasing the mining time.
In \cite{lim03}, a technique for compressing large transactions is
proposed --  however, its utility is primarily in reducing storage and
communication cost, and not in reducing the mining runtime since the
compressed transactions have to be decompressed during the mining process.

\item[Counting Overhead:]~
On distortion, a $k$-itemset may produce any of $2^k$ combinations.
For example, a '11' may distort to '00','01','10' or '11'. In
order to accurately reconstruct the support of the $k$-itemset, we need
to, in principle, keep track of the counts of all $2^k$ combinations.
To reduce these costs, MASK took the approach of maintaining
\emph{equal} flipping probabilities for both 1's and 0's -- with this
assumption, the number of counters required is only $k$~\cite{mask}.
Further, only $k-1$ of the counts need to be explicitly maintained since
the sum of the counts is equal to the database cardinality, $dbsize$ --
the counter chosen to be implicitly counted was the one with the
expected largest count.

While the above counting optimizations did appreciably reduce runtime costs, 
yet the overhead in absolute terms remains very significant -- as mentioned
earlier, it could be as much as two to three orders of magnitude compared to
the time taken for direct mining.
\end{description}

\noindent
In the remainder of this section, we describe how EMASK is engineered
to address the above efficiency concerns.

\comment{
The reason for MASK's huge overheads are the following:  

\begin{description}
\item[Support reconstruction overhead:]~
On randomization a k-itemset may produce a subset of length 0 to k. A '11'
may distort to '00','01','10' or '11'. To reconstruct the support of the
original '11' we need to keep track of all these combinations to obtain k
counts required for constructing the distorted matrix. This extra counting
overhead poses significant increase in mining time. We show in Section
\ref{sec:Elimnating counting overhead} a method to eliminate this overhead
completely.

\item[Long transactions:]~
By applying randomization to generate false items, we increase the length
of transaction largely. eg Randomizing a transaction by probability p=0.9 
increased the average transaction length from 10 to 110. Hence to
count the itemsets in the distorted database requires much more processing
than the original database. This increases the mining time largely.
\cite{lim} looks into the problem of large transactions and proposes a
method of compressing the large transactions in order to reduce storage
and communication cost. But compression does not help in reducing the
mining time. Transactions need to be decompressed to mine association
rules.

\end{description}
}

\subsection{The Distortion Process}
\label{sec:distort}
The new feature of EMASK's distortion process is that it applies
\emph{symbol-specific} distortion -- that is, 1's are flipped with
probability $(1-p)$, while 0's are flipped with probability $(1-q)$.
Note that in this framework the original MASK can be viewed as a special
case of EMASK wherein $p = q$.

The idea here is that MASK generates false items to hide the true
items that are retained after distortion, resulting in an increase in
database density. But, if a fewer number of true items are retained, a
fewer number of false items need to be generated, and we can minimize
this density increase.  Thus the goal of reduced density could be
achieved by reducing the value of $p$ and increasing the value of $q$
(specifically increasing it to beyond 0.9).  However, note that a
decrease in $p$ value increases the distortion significantly which
can reduce accuracy of reconstruction. Also, $q$ or the non-flipping
probability of 0s cannot be increased to very high values as it can
decrease the privacy significantly.  Thus the choices of $p$ and $q$
have to be made \emph{carefully} to obtain a combination of $p$ and $q$
values, such that $q$ is high enough to result in decreased transaction lengths
but privacy and accuracy are still achievable.

We defer the discussion
on how to select appropriate values of $p$ and $q$ to Section~\ref{sec:pqselection}.

\comment{
To select the distortion parameters we use the observation that the
increase in length of randomized transactions is due to large number of
false items generated, i.e. a large number of 0s being turned to 1. On
the other hand generation of false items is basic for privacy. False
items are generated to hide the true items which are retained after
randomization. Hence increasing $q$ to decrease false items may
adversely effect privacy.  But, if less number of true items are
retained, we may be able to get same privacy even on generating less
false items, i.e. same level of privacy may be obtained for higher
values of $q$ if $p$ is decreased.  Hence we try to observe the region
with $q>0.9$ and $p<0.9$ to meet the three goals of privacy, accuracy
and efficiency.
}



\subsubsection{Privacy Estimation for EMASK}
\label{sec:priv}
As mentioned earlier, privacy can be computed at two levels: Basic
Privacy (BP) and Reinterrogated Privacy (RP). We now quantify
the privacy provided by EMASK with regard to both these metrics.

\subsubsection{Basic Privacy}
\label{sec:baspriv}
The basic privacy measures the probability that given a random customer
$C$ who bought an item $i$, her original entry for $i$, i.e. '1',
can be accurately reconstructed from the distorted entry, \emph{prior}
to the mining process.
We can calculate this privacy in the following manner, similar to the
procedure outlined for MASK in \cite{mask}: Let $s_i$ be the support of
the $i^{th}$ item, $X_i$ be the original entry corresponding to item
$i$, in a tuple and $Y_i$ be its distorted entry.  The reconstruction
probability of a $1$ is given by
\[
R_1(p,q,s_i) = \Pr(Y_i=1 | X_i = 1)  \Pr(X_i=1 | Y_i = 1)
                 + \Pr(Y_i=0 | X_i = 1)  \Pr(X_i=1 | Y_i = 0)
\]
\[  = \frac{ {\Pr(Y_i=1 | X_i = 1)}^2  \Pr(X_i=1) } { \Pr(Y_i = 1)}
    + \frac{ {\Pr(Y_i=0 | X_i = 1)}^2  \Pr(X_i=1) } { \Pr(Y_i = 0)}
\]
\[   = \frac{ p^2  s_i } { \Pr(Y_i = 1)}
    + \frac{ (1-p)^2  s_i } { \Pr(Y_i = 0)}
\]
But,
\[ \Pr(Y_i = 1) = \Pr(Y_i = 1| X_i = 0) \Pr(X_i = 0) + \Pr(Y_i=1 | X_i = 1) \Pr(X_i=1)
\]
\[ = s_i p + (1-s_i)(1-q)
\]
And,
\[ \Pr(Y_i = 0) = 1 - \Pr(Y_i = 1) = s_i (1-p) + (1-s_i) q
\]
Therefore, 
\[R_1(p,q,s_i) = \frac{ p^2  s_i } {s_i p + (1-s_i)(1-q) }
    + \frac{ (1-p)^2  s_i } {s_i (1-p) + (1-s_i) q }
\]
The overall reconstruction probability of 1's is now given by
\[
 R_1(p,q) = \frac{\sum_{i} s_i R_1(p,q,s_i)} {\sum_{i} s_i}
\]
and the Basic privacy offered to 1's is simply
$100(1 - R_1(p,q))$.

The above expression is \emph{minimized} when all the items in the
database have the same support, and increases with the variance in the
supports across items.  As discussed later in Section~\ref{sec:expt},
with the appropriate choices of $p$ and $q$, this increase is marginal for the
market-basket type of datasets that we consider here.  Therefore, as a
first-level approximation, we replace the item-specific supports in the
above equation by $s_{0}$, the average support of an item in the database.
With this, the reconstruction probability simplifies to

\begin{equation}
\textstyle
R_1(p,q) = \frac{ p^2  s_0 } {s_0 p + (1-s_0)(1-q) }
    + \frac{ (1-p)^2  s_0 } {s_0 (1-p) + (1-s_0) q }
\label{eq:recon-emask}
\end{equation}

\begin{figure}[ht] 
\begin{center}
\includegraphics[width=0.4\textwidth]{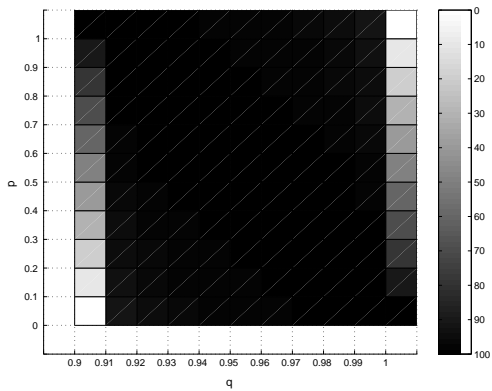}
\caption{Basic Privacy of EMASK
\label{fig:privacyplot}}
\end{center}
\end{figure}

In Figure~\ref{fig:privacyplot} we plot $100(1-R_1(p,q))$ i.e. privacy for
different values of $p$ and $q$ at $s_0$=0.01 -- the shadings indicate
the privacy ranges.  Note that the color of each square represents the
privacy value of the lower left corner of the square.  We observe here
that there is a \emph{band} of values for $p$ and $q$ in which the privacy
is greater than the 80\% mark. Specifically, for $q$ between $0.1$ and
$0.9$, high privacy values are attainable for all $p$ values, and the
whole of this region appears black.  Beyond $q=0.9$, privacy above 80\%
is attainable but only for low $p$ values.  Similarly for $q$ below 0.1,
high privacy can be obtained only at high $p$ values.

\subsubsection{Reinterrogation privacy}
\label{sec:repriv}
Reinterrogation privacy takes into account the reduction in privacy
due to the knowledge of the \emph{output} of the mining process --
namely, the association rules, or equivalently, the support of frequent
itemsets~\cite{as94}.  Privacy breach due to reinterrogation stems
from the fact that an individual entry in the distorted database may
not reveal enough information to reconstruct it, but on seeing a long
frequent itemset in the distorted database and knowing the distorted
and original supports of the itemset one may be able to predict the
presence of an item of the itemset with more certainty.  As an example
situation, suppose the reconstructed support of a 3-itemset present in a
transaction distorted with $p=0.9, q=0.9$, is 0.01. Then the probabiity
of this 3-itemset coming from a '000' in the original transaction is as
low as $0.1*0.1*0.1*0.99=0.00099$. Thus, with the knowledge of support
of higher length itemsets the miner can predict the presence of an item
of the itemset in the original transaction with higher probability.

Our method of estimating reinterrogation privacy breach is based on that
described in \cite{gehrke02} for computing privacy breaches.  We make the
\emph{extremely conservative} assumption here that the data miner is able
to accurately estimate the \emph{exact} supports of the original database
during the mining process.  The method of calculating reinterrogation
privacy is as follows: \comment{ We use the original and randomized
databases as well as the list of frequent itemsets to estimate average
reinterrogation privacy obtained by a given database. Our estimation
procedure is as follows.  }

\jhcomment{There is also an assumption here that the miner is using
only the support and confidence levels specified by the user -- what
if he uses a very low support level - almost all itemsets will be
frequent and the privacy will go down.}

\begin{itemize}
 
\item First, for each item that is part of a frequent itemset, we compare
the support of the frequent itemset in the distorted data with the
support of the singleton item in the original data.  For example, let an
itemset \{a,b,c\} occur a hundred times in the randomized data and let
the support of the item $b$ in the corresponding hundred transactions of
the original database be twenty.  We then say that the itemset \{a,b,c\}
caused a 20\% privacy breach for item $b$ due to reinterrogation,
since for these hundred distorted transactions, we estimate with 20\%
confidence that the item $b$ was present in the original transaction.

\item Then, we estimate the privacy of each '1' appearing in a frequent
item column in the original database. There are two cases: 1's that
have remained 1's in the distorted database, and 1's that have been
flipped to 0's. For the former, the privacy is estimated 
with regard to the worst of the privacy breaches (computed as
discussed above) among all the frequent itemsets of which it is a part and
which appear in the distorted version of this transaction. For example,
for an item $b$ in the original transaction \{a,b,c,d,e\},
the privacy breach of $b$ is the worst of the privacy breaches due to
all frequent itemsets which are subsets of \{a,b,c,d,e\} and contain $b$.

In the latter (flip) case, the privacy is set equal to the average Basic
Privacy value -- this is because for the $q$ values that we consider
in this paper, which are close to 1.0 as discussed later, most of the
large number of $0$'s in the original sparse matrix remain $0$, so a $1$
flipping to a $0$ is resistant to privacy breaches due to reinterrogation.

\item Next, for the 1's present in in-frequent columns,
the privacy is estimated to be the Basic Privacy since these
privacies are not affected by the discovery of frequent itemsets.

\item Finally, the above privacy estimates are averaged over all
1's in the original database to obtain the overall average 
reinterrogation privacy.

\end{itemize}

\comment{

We consider the effect of frequent itemsets (for a sufficientely low $sup_{min}$) 
alone in the reduction of privacy. 
The reasons for this choice are : 
\begin{itemize}
\item Computing privacy breaches taking all itemsets into account is
computationally infeasible  in practice for long transactions. For
instance, in a transaction of length = 40, there are $2^{40} \simeq
10^{12}$ itemsets which are a subset of the transaction.
\item The supports of itemsets that are very sparse are difficult to
compute with a good accuracy by reconstruction.
\item It is the frequent itemsets that give the most relevant information
that could contribute to privacy breaches.
\end{itemize}

Basically if we keep reducing the minsup value, and thereby include more itemsets to 
consider breaches of, the additional gain of the low support itemsets in reducing privacy
 becomes less and less(also the percentage error in low supports will generally be higher). 
Thus the privacy wouldn't drop substantially even if all itemsets were to be considered.

}

\subsection{The EMASK Mining Process}
\label{sec:mine}
Having established the privacy obtained by EMASK's distortion process,
we now move on to presenting EMASK's technique for estimating the
true supports of itemsets from the distorted database.
In the following discussion, we first show how to estimate the supports
of 1-itemsets (i.e. singletons) and then present the general $n$-itemset
support estimation procedure.  In this derivation, it is important to
keep in mind that the miner is provided with both the distorted matrix
as well as the distortion procedure, that is, he \emph{knows} the values
of $p$ and $q$ that were used in distorting the true matrix.

\subsubsection{Estimating Singleton supports}
We denote the original true matrix by $T$ and the distorted
matrix, obtained with distortion parameters $p$ and $q$, as $D$.
Now consider a random individual item $\bf{i}$.  Let $c^T_1$ and $c^T_0$
represent the number of 1's and 0's, respectively, in the $\bf{i}$
column of ${T}$, while $c^D_{1}$ and $c^D_{0}$ represent the number
of 1's and 0's, respectively, in the $\bf{i}$ column of ${D}$.
With this notation, we estimate the support of $\bf{i}$ in $T$
using the following equation:
 
\begin{equation}
\bf{C^T=M^{-1}C^D}
\label{eq:matrix}
\end{equation}
 
\noindent
where
\[
M=\left[\begin{array}{cc}
p & 1-q \\
1-p & q
\end{array}\right]
\;
C^D=\left[\begin{array}{c}
c^D_{1} \\
c^D_{0}
\end{array}\right]
\;
C^T=\left[\begin{array}{c}
c^T_{1} \\
c^T_{0}
\end{array}\right]
\]


\subsubsection{Estimating $n$-itemset Supports}
It is easy to extend Equation~\ref{eq:matrix}, which is applicable to
individual items, to compute the support for an arbitrary $n$-itemset. For
this general case, we define the matrices as:
 
\[
C^D=\left[\begin{array}{c}
c^D_{2^{n}-1} \\
. \\
. \\
. \\
c^D_{1}\\
c^D_{0}
\end{array}\right]
\; \; \;
C^T=\left[\begin{array}{c}
c^T_{2^{n}-1} \\
. \\
. \\
. \\
c^T_{1}\\
c^T_{0}
\end{array}\right]
\]
 
\noindent 
Here $c^T_{k}$ should be interpreted as the count of the tuples in ${T}$
that have the binary form of $k$ (in $n$ digits) for the given itemset
(that is, for a 2-itemset, $c^T_{2}$ refers to the count of 10's in the
columns of ${T}$ corresponding to that itemset, $c^T_{3}$ to the count
of 11's, and so on).  Similarly, $c^D_{k}$ is defined for the distorted
matrix ${D}$.  

The column matrices can be simplified by observing that the distortion
of an entry in the above distortion procedure depends only on whether
the entry is 0 or 1, and \emph{not} on the column to which the entry
belongs, rendering distortion of all combinations with same number of
1s and 0s equivalent.  Hence the above matrices can be represented as:

\[ C^D=\left[\begin{array}{c} c^D_{n} \\ . \\ . \\ .\\ c^D_{1} \\
c^D_{0} \end{array}\right] \; \; \; C^T=\left[\begin{array}{c} c^T_{n}
\\ . \\ . \\ . \\ c^T_{1}\\ c^T_{0} \end{array}\right] \]

\noindent
where $c^T_{k}$ should be interpreted as the count of the tuples in
${T}$ that have the binary form with $k$ 1's (in $n$ digits) for the given
itemset. For example , for a 2-itemset, $c^T_{2}$ refers to the count of
11's in the columns of ${T}$ corresponding to that itemset, $c^T_{1}$ to
the count of 10's and 01's, and $c^T_{0}$ to the count of 00's.  Similarly,
$c^D_{k}$ is defined for the distorted matrix ${D}$.

Each entry $m_{i,j}$ in the matrix $\bf{M}$ is the probability
that a tuple of the form corresponding to $c^T_{j}$ in ${T}$ goes to a
tuple of the form corresponding to $c^D_{i}$ in ${D}$.
For example, $m_{2,1}$ for a 2-itemset is the probability that a $10$ or
$01$ tuple distorts to a $11$ tuple. Accordingly, $m_{2,1}$ = $p(1-q)$.
The basis for this formulation lies in the fact that in our distortion
procedure, the component columns of an $n$-itemset are distorted
\emph{independently}. Therefore, we can use the product of the
probability terms.  In general,
\begin{equation}
m_{i,j}=\sum_{k=max(0,i+j-n)}^{min(i,j)} \choose{j}{k} \; p^k (1-p)^{(i-k)} \; \; \choose {n-j}{i-k} \; q^{(n+k-i-j)} (1-q)^{(i-k)}
\label{eq:matent}
\end{equation}



                                                    

\subsection{Eliminating Counting Overhead}
\label{sec:opt}

We now present a simple but powerful optimization by which the entire 
extra overhead of counting all the combinations generated by the
distortion can be \emph{eliminated completely}.  This optimization is
based on the following basic formula from set theory:  Given a universe $U$,
and subsets $A$ and $B$,
        \[ N(A' \cap B)=N(B) - N(A \cap B) \]
\noindent
where N is the number of elements, i.e. cardinality, of the set denoted
by the bracketed expression.  This formula can be generalized\footnote{
$ N(B_1 B_2 ..B_n) $ is replaced by $ N(U) $ if $ n=0 $} to \\
\[
N(A_1' A_2' ...A_m' B_1 B_2 ...B_n)
\]
\[
= N(B_1 B_2 ..B_n) \,+\, \sum _{k=1} ^m \sum _{\{x_1,...,x_k\} \subset
\{1,...,m \} } (-1)^k N(A_{x_1} A_{x_2} ... A_{x_k} B_1 B_2 ..B_n)  
\]

\noindent
Using above formula the counts of all the combinations generated from an
$n$-itemset can be calculated from the counts of itemsets and the counts
of their subsets which are available from previous passes over the
distorted database. \emph{Hence, we only need to explicitly count 
only the '111...1's, just as in the direct mining case}.

For example, during the second pass we need to explicitly count only
'11's which makes $N(A \cap B )$ available at the end of the second pass.
The counts of the remaining combinations, '10', '01' and '00' can then
be calculated using the following set of formulae: \\
\noindent
\hspace*{0.5in} 10 : $ N(A \cap B') = N(A) - N(A \cap B)$ \\ 
\hspace*{0.5in} 01 : $ N(A' \cap B) = N(B) - N(A \cap B) $ \\
\hspace*{0.5in} 00 : $ N(A' \cap B')= N(D) - N(A) - N(B) + N(A \cap B) $

\noindent
The above implies that the extra calculations for reconstruction are
performed only at \emph{the end of each pass}, the rest of the pass being
exactly the same as that of the original mining algorithm. Further,
the only additional requirement of the above approach as compared to traditional
data mining algorithms such as Apriori is that we need to have available,
at the end of each pass, the counts of all frequent itemsets generated
during the previous passes. However, this requirement is easily met by
keeping a hash table in memory of these previously identified frequent
itemsets.

\comment{
This reduction in counting had significant contribution in decreasing
running time.

Note that here we assume that we have the counts of large itemsets from
all the previous passes available at each pass. Apriori doesn't require
the counts from the previous passes in any pass. But this is not much
cost. They can be either kept into a hash table into memory or may be
read from disk if required, which shall be only once at the end of a pass.
}

\subsection{Choosing values of $p$ and $q$}
\label{sec:pqselection}

As promised earlier, we now discuss how the distortion parameters, $p$
and $q$, should be chosen. Our goal here is to identify those parameter settings 
that simultaneously ensure acceptable levels of both privacy and accuracy. 
We use privacy and accuracy estimations to choose these values. The average 
support value of the dataset are required for these estimations. We assume here that some 
initial sample data is available based on which the determination of average support value 
can be made.

 Earlier in this
section, we had shown how the basic privacy can be calculated. We use the basic privacy 
values as the estimation for privacy achievable at a point ($p$,$q$) as the basic privacy can be
 calculated before distortion.
We now describe how to estimate the accuracy of reconstruction beforehand.  
Our focus is specifically on
$1$-itemsets since their error percolates through the entire mining process.

Consider a single column in the true database which has $\it{n}$ 1's
and $\it{dbsize-n}$ 0's. We expect that the $\it{n}$ 1's will distort
to $\it{np}$ 1's and $\it{n(1-p)}$ 0's when distorted with parameter
$\it{p}$. Similarly, we expect the 0's to go to $\it{(dbsize-n)q}$ 0's
and $\it{(dbsize-n)(1-q)}$ 1's. However, note that this is assuming that
``When we generate a random number, which is distributed as bernoulli(p),
then the number of 1's, denoted by P, in $\it{n}$ trials is actually
$\it{np}$''. But, in reality, this will not be so.  Actually, P is
distributed as \emph{binomial(n,p)}, with mean $np$ and variance
$np(1-p)$.

\noindent

The total number of 1's in a column of the distorted matrix
can be expressed as the sum of two random variables, $X$ and $Y$:
\begin{itemize}
\item $X$: The 1's in the distorted database that come from 1's in the original
database.
\item $Y$: The 1's in the distorted database that come from 0's in the original
database.
\end{itemize}
The variance in the total number of 1's generated in the distorted
matrix is $var(X+Y)$, which is simply $var(X)+var(Y)$ since $X$
and $Y$ are \emph{independent} random variables. So, we have $var(X+Y) =
var(X)+var(Y) = np(1-p) + (dbsize-n)(1-q)q$.  Therefore, the deviation
of the number of 1's in the distorted database from the expected value
is 
\begin{equation}
{\triangle n'} = (np(1-p) + (dbsize-n)(1-q)q )^{1/2}
\label{eq:stddev}
\end{equation}

Let the distorted column have $\it{n'}$ 1's. Then, our estimate of
the original count is given by
\[
\overline{n} = \frac{n'}{p+q-1} - \frac{(1-q)dbsize}{p+q-1}
\label{eq:recons}
\]
and the possible error in this estimation is computed as
\[
\triangle \overline{n} = \frac{\triangle n'}{p+q-1}
\label{eq:error1}
\]
where  ${\triangle n'}$ is as per Equation~\ref{eq:stddev}. If we normalize
the error to the true value ($n$), we obtain
\begin{equation}
Error = \frac{(p(1-p) + ({\frac{dbsize}{n}} - 1)(1-q)q )^{1/2}}{p+q-1}
\label{eq:error2}
\end{equation}
which is completely expressed in terms of known parameters.

Using Equations~\ref{eq:recon-emask} and ~\ref{eq:error2}, we can set
the $p$ and $q$ parameters so as to achieve reasonable 
privacy and accuracy goals.

\section{Performance Framework}
\label{sec:perf}
EMASK aims at simultaneously achieving satisfactory performance on three
objectives: privacy, accuracy and efficiency. While privacy is determined
as outlined in Section~\ref{sec:priv}, the specific metrics used
to evaluate EMASK's performance w.r.t. accuracy and efficiency are given
below.

\subsection{Accuracy Metrics}
\label{sec:errormetrics}
\comment{
Since EMASK is a \emph{probabilistic} approach, fundamentally we cannot
expect the reconstructed support values to coincide exactly with
the actual supports.  This means that we may have errors in the
estimated supports of frequent itemsets with the reported values
being either larger or smaller than the actual supports. 

Errors in support estimation can have an even more pernicious effect
than just wrongly reporting the support of a frequent itemset. They
Performance can result in errors in the \emph{identities} of the frequent itemsets.
This becomes especially an issue when the $sup_{min}$ setting is such that
the support of a number of itemsets lie very close to this threshold
value. Typically, this happens for lower values of $sup_{min}$ due to the
larger number of frequent itemsets in the database.  Such ``border-line''
itemsets may get wrongly reported as either frequent or rare, based on
how the probabilistic evaluation estimates their supports.  That is,
we can encounter both \emph{false positives} and \emph{false negatives}.

Worse, errors in association rule mining \emph{percolate} through the
various passes of the mining process -- that is, an error in identifying
a 1-itemset correctly has a ripple effect in terms of causing errors
in the remainder of the frequent itemset lattice.
}

We evaluate two kinds of mining errors, Support Error and Identity Error,
in our experiments: 
\begin{description}
\item[Support Error ($\rho$)]:~\\
This metric reflects the (percentage) average relative error in the
reconstructed support values for those itemsets that are correctly
identified to be frequent. Denoting the number of frequent itemsets by $|f|$, 
the reconstructed support by
$rec\_sup$ and the actual support by $act\_sup$, the support error
is computed over all frequent itemsets as
\[
{\rho}  = \frac{1}{\mid f \mid}  \Sigma_{f} 
\frac{\mid rec\_sup_f - act\_sup_f \mid}{act\_sup_f} * 100
\]


\item[Identity Error ($\sigma$)]:~\\
This metric reflects the percentage error in identifying frequent itemsets
and has two components: $\sigma^+$, indicating the percentage of false
positives, and $\sigma^-$ indicating the percentage of false
negatives.  Denoting the reconstructed set of frequent itemsets with $R$
and the correct set of frequent itemsets with $F$, these metrics are
computed as:
\begin{center}
${\sigma^+}  = \frac{\mid R - F\mid }{\mid F\mid } * 100$ \hspace*{0.5in}
${\sigma^-}  = \frac{\mid F - R\mid }{\mid F\mid }$ * 100
\end{center}
\end{description}


\subsection{Efficiency Metric}
\label{sec:effmetrics}
This metric determines the runtime overheads resulting from mining the
distorted database as compared to the time taken to mine the original
database. This is simply measured as the inverse ratio of the running
times between Apriori on the original database and executing the same
code augmented with EMASK (i.e. EMASK-Apriori) on the distorted database.  
Denoting this slowdown ratio as ${\Delta}$, we have
\begin{center}
${\Delta}  = \frac{RunTime \;  of \; EMASK\-Apriori}{RunTime \;  of \; Apriori}$ 
\end{center}
For ease of presentation, we hereafter refer to
this augmented algorithm simply as EMASK.

\subsection{Data Sets}
We carried out experiments on a variety of synthetic and real datasets. Due to space
limitations, we report the results for only two representative databases here:
\begin{enumerate}
\item
A synthetic database generated from the IBM Almaden
generator~\cite{as94}. The synthetic database was created with
parameters T10.I4.D1M.N1K (as per the naming convention
of \cite{as94}), resulting in a million customer tuples
with each customer purchasing about ten items on average.
\comment{The number of patterns was 2000.}

\item
A real dataset, BMS-WebView-1~\cite{real}, placed in the public domain
by Blue Martini Software. This database contains click-stream data from the
web site of a (now defunct) legwear and legcare retailer.  There are
about 60,000 tuples with close to 500 items in the schema. In order to
ensure that our results were applicable to large disk-resident databases,
we scaled this database by a factor of ten, resulting in approximately
0.6 million tuples.

\comment{
\item
Another real dataset, BMS-WebView-2~\cite{real}, placed in public domain
by Blue Martini Software. This dataset also contains click-stream data
from e-commerce websites. There are about 78,000 tuples in the database
with 3340 items in the schema. This database was scaled by 15 times
resulting in approximately 1 million tuples. The average transaction
length was 5 for this database.
}

\end{enumerate}

\subsection{Support and Distortion settings}
\label{sec:supsettings}

\begin{figure}
\begin{center}
\subfigure[] {\includegraphics[width=0.4\textwidth]{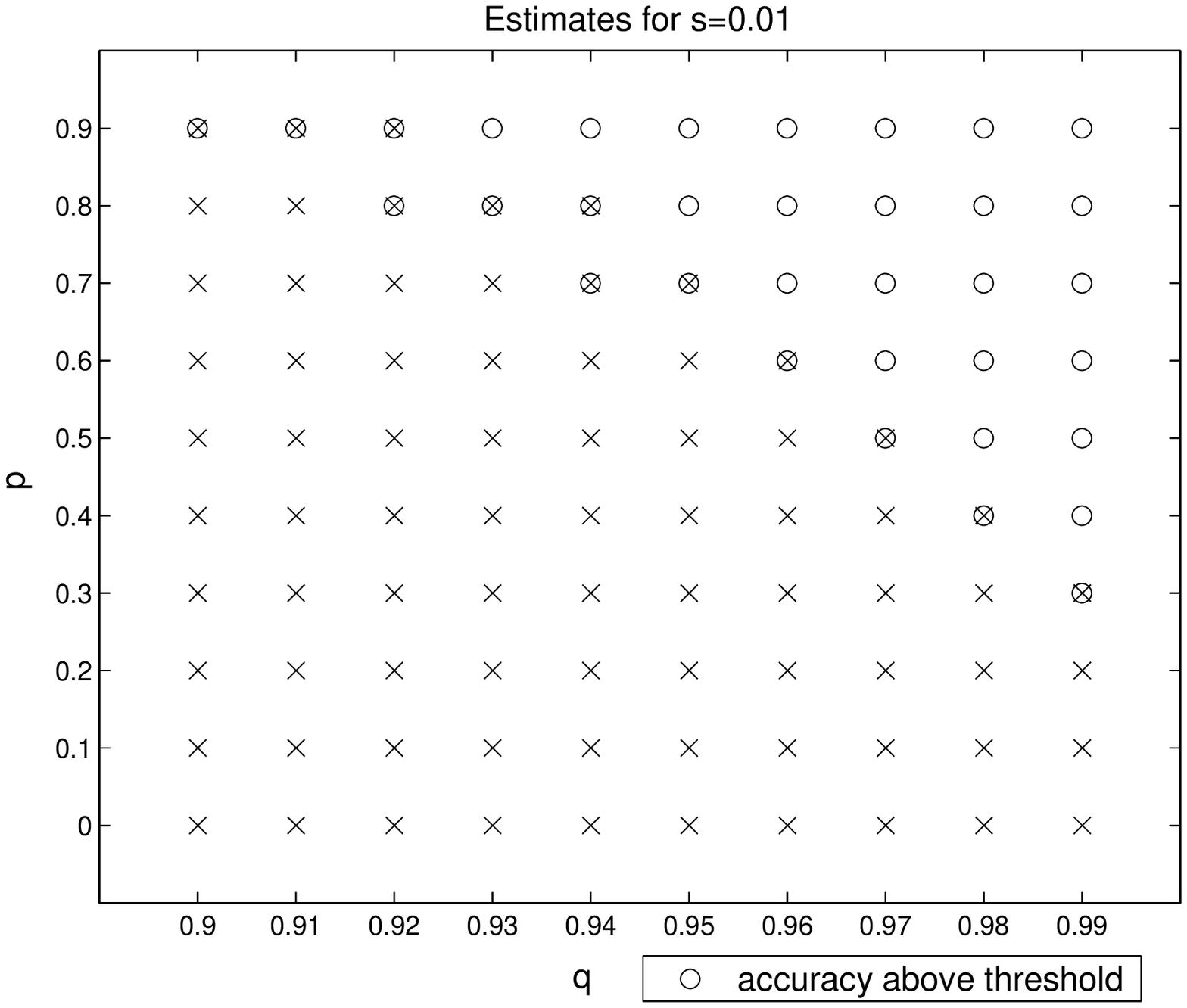}}
\hspace{0.5cm} 
\subfigure[] {\includegraphics[width=0.4\textwidth]{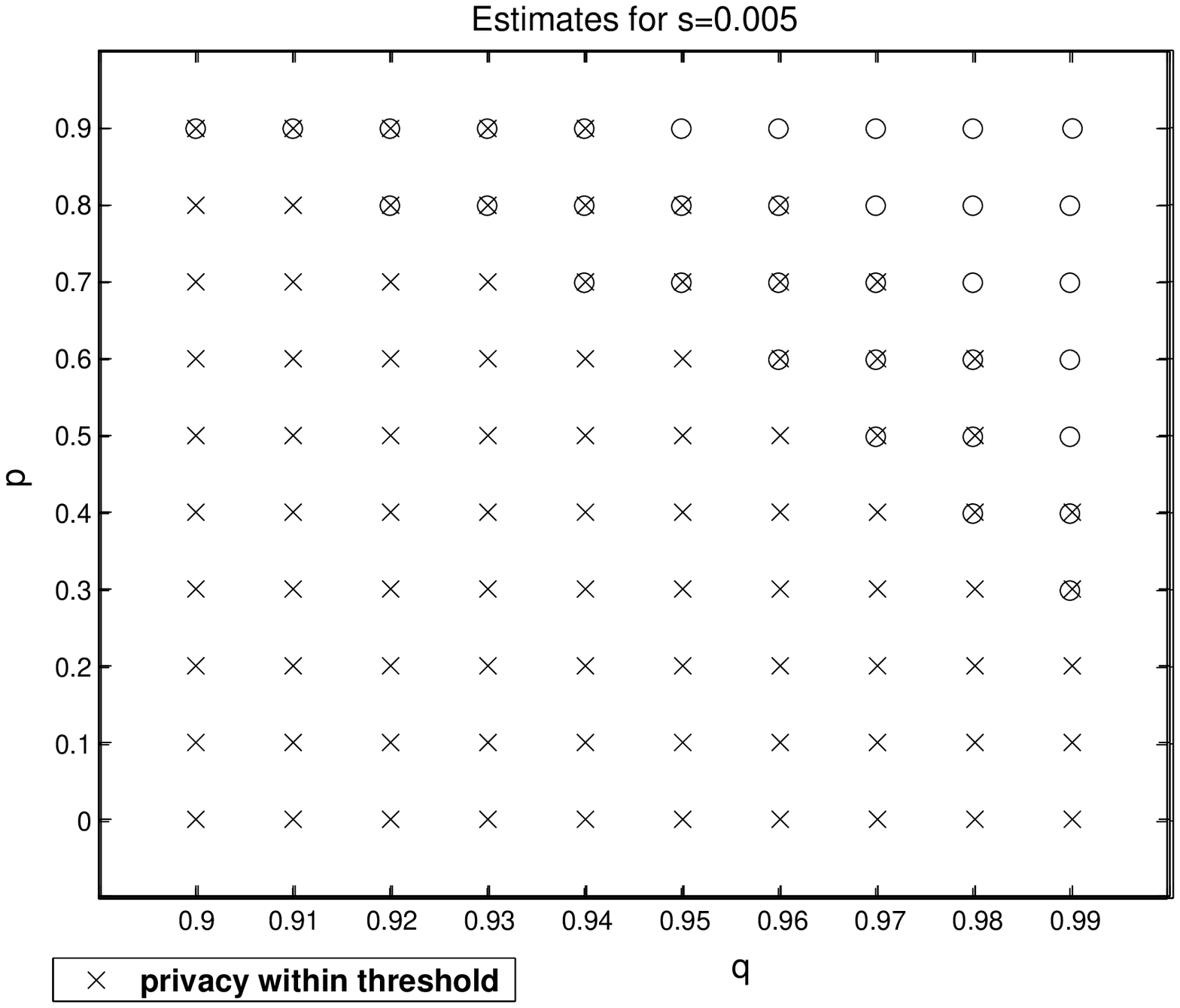}} 
\caption{Determining Acceptable $p,q$ Candidates
\label{fig:des}} 
\end{center}
\end{figure}

The theoretical basis for determining the settings of the distortion
parameters, $p$ and $q$,  was presented in Section~\ref{sec:pqselection}.
We now utilize those formulas to derive acceptable choices for the
datasets mentioned above.  The privacy and accuracy estimates for $p$
values ranging from 0.1 to 0.9 and $q$ values ranging from 0.9 to 0.99 are
shown in Figures~\ref{fig:des}(a) and ~\ref{fig:des}(b) for support=0.01
(the synthetic dataset) and support=0.005 (real dataset), respectively.
In these figures, the symbol {\tt o} represents $p,q$ combinations for
which the basic privacy is above a minimum threshold value, while the
symbol {\tt x} represents combinations for which the accuracy is above
a minimum threshold value.  The threshold for basic privacy was set at
90\% while the threshold chosen for accuracy was that produced by the
original MASK algorithm, that is, the accuracy estimate obtained with
$p=0.9,q=0.9$\footnote{Choosing this threshold instead of a fixed
value allows adaptation to the characteristics of the specific data
set that is being mined.}.

The points in Figures~\ref{fig:des}(a) and ~\ref{fig:des}(b) which
have \emph{both symbols} (i.e. {\tt o} and {\tt x}) represent $p,q$
combinations that have both good privacy and accuracy. For example,
$p=0.5,q=0.97$ is such a combination in both figures. From among this
set of points, we select only those combinations in which $q$ is greater
than 0.95 -- this is because, as mentioned earlier, from an efficiency
perspective we would like to have $q$ as high as possible since it
directly impacts the density of the distorted database.

With the above considerations in mind, the shortlist of candidate
combinations is shown in Tables~\ref{tab:pq_syn} and \ref{tab:pq_bms1}
for synthetic and real datasets, respectively.  The fact that the
real dataset has a richer set of candidates than the synthetic dataset
is in accordance with the observation in \cite{mask} that the sparser
the dataset, the more amenable it is to privacy-preserving mining.

\begin{table}
\small
\begin{center}
\begin{tabular}{|c|c|}
\hline
\emph{p} & \emph{q} \\
\hline
0.3 & 0.99 \\
\hline
0.4 & 0.98 \\
\hline
0.5 & 0.97 \\
\hline
0.6 & 0.96 \\
\hline

\end{tabular}
\caption{Candidate $p,q$ combinations for synthetic database
\label{tab:pq_syn}}
\end{center}
\end{table}

\begin{table}
\small
\begin{center}
\begin{tabular}{|c|c|c|c|}
\hline
\emph{p} & \emph{q} & \emph{p} & \emph{q} \\
\hline
0.3 & 0.99 & 0.6 & 0.98 \\
\hline
0.4 & 0.99 & 0.6 & 0.97 \\
\hline
0.4 & 0.98 & 0.6 & 0.96 \\
\hline
0.5 & 0.98 & 0.7 & 0.97 \\
\hline
0.5 & 0.97 & 0.7 & 0.96 \\
\hline

\end{tabular}
\caption{Candidate $p,q$ combinations for real database
\label{tab:pq_bms1}}
\end{center}
\end{table}

\jhcomment{It would be good to put the tables side-by-side}
 
\comment{
The 0.3\% $sup_min$ value represents, in a sense, the ``worst case"
environment for our algorithm due to the presence of a large number of
border-line itemsets.
}

\shcomment{label the figures separately}

\section{Experimental Results}
\label{sec:expt}
We evaluated the privacy, accuracy and efficiency of the EMASK privacy-preserving
mining process for the candidate distortion parameters on the two datasets for a variety
of minimum support values. Due to space limitations, we present here
the results only for 0.3\% $sup_{min}$ value, which represents a support
low enough for a large number of frequent itemsets to be produced,
thereby stressing the performance of the EMASK algorithm.  The results
for the synthetic database are presented first, followed by those for
the real database.

\shcomment{In what order p \& q result should be writen}

\subsection{Experiment Set 1 : Synthetic Dataset}
Mining the synthetic dataset with a $sup_{min}$ of 0.3 resulted in frequent
itemsets of length upto 8.  Table \ref{tab:syn_results} presents the EMASK
privacy, accuracy and efficiency results for this dataset in a summarized
fashion. In this table, for each $p,q$ candidate, \emph{BP} and \emph{RP}
denote the basic and re-interrogated privacies, respectively; $\Delta$
denotes the slowdown of EMASK; and the remaining columns show the
accuracy metrics. We also include the variances (across mining levels)
of the accuracy metrics to determine whether the accuracy was skewed
based on the lengths of the frequent itemsets.

\begin{table}[ht]
\small
\begin{center}
\begin{tabular}{|c|c|c|c|c|c|c|c|c|c|c|}
\hline
p & q & BP & RP & $\Delta$ & $\sigma^+$ & $\sigma^-$ & $\rho$ & 
$Var(\sigma^+)$ & $Var(\sigma^-)$ & $Var(\rho)$ \\
\hline
0.6 & 0.96 & 92 & 70 & 5.2 & 4.99 & 5.34 & 3.39 & 3.23 & 6.84 & 0.27 \\
\hline
0.5 & 0.97 & 92.6 & 74 & 3.8 & 5.64 & 6.27 & 4.86 & 7.74 & 6.41 & 4.31 \\
\hline
0.4 & 0.98 & 93 & 78 & 2.4 & 6.40 & 7.87 & 6.60 & 6.13 & 18.35 & 19.02 \\
\hline 
0.3 & 0.99 & 92.6 & 80 & 1.1 & 6.63 & 11.69 & 10.19 & 7.61 & 137.20 & 120.24\\
\hline
\end{tabular}
\caption{Performance Results for Synthetic dataset
\label{tab:syn_results}}
\end{center}
\end{table}

\comment{
The subsequent columns indicate the error in reconsruction as per
the accuracy metrics.  It gives for all the chosen p \& q values,
the average error in terms of average $\sigma^+$, average $\sigma^-$
and average $\rho$ values accross various levels. Also the variance in
$\sigma^+$,$\sigma^-$ \& $\rho$ across levels is shown.
}

\shcomment{Need to use some symbols for average and variance , may be sigmaplus bar,
'rho bar 'and so on} 

The first point to note in Table \ref{tab:syn_results} is that the results
are very encouraging since they clearly demonstrate that EMASK is able to
get high values for all three competing objectives -- for example,
with $(0.4,0.98)$, the basic and re-interrogated privacy are above 75
percent, the slowdown is only 2.4 (that is, it performed only around twice
as slow as Apriori), and the accuracy on all measures is better than 90
percent (less than 10 percent error).

Secondly, note that the $\Delta$ slowdown values across all the
combinations is such that EMASK always performs \emph{within 5 times of Apriori}.
This is indeed a \emph{huge improvement} from the several orders of
magnitude inefficiency that was exhibited by MASK, as discussed earlier
in Section~\ref{sec:emask}.

Third, while it may appear that $(0.3,0.99)$ is very desirable from
privacy and efficiency perspectives (it executes almost as quickly
as Apriori and has an 80-plus privacy), yet it may not be a suitable
choice since the \emph{variance} in the accuracy measures is very high.
Specifically, the longer (and potentially most important) frequent
itemsets have much higher errors with the combinations with low $p$ values like this.
Table \ref{tab:06096} and Table \ref{tab:03099} show errors at each level
for ${0.6,0.96}$ (highest $p$ value in the candidates) and ${0.3,0.99}$
(lowest $p$ value) respectively.  In the tables, the Level indicates the
length of the frequent itemsets, $|F|$ indicates the number of frequent
itemset at this level, and other three columns are error metrics as
discussed before.  The tables indicate that  to ensure high accuracy
at higher levels one must move to high $p$ values among the candidate
combinations.

\begin{table}[ht]
\small
\begin{center}
\begin{tabular}{|c|c|c|c|c|}
\hline
Level & $|F|$ & $\sigma^+$ & $\sigma^-$ & $\rho$ \\
\hline
1 & 664 & 1.5 & 1.5 & 3.21 \\
\hline
2 & 1847 & 5.79 & 5.46 & 3.38 \\
\hline
3 & 1310 & 3.66 & 4.27 & 2.9 \\
\hline
4 & 864 & 7.29 & 6.13 & 3.45 \\
\hline
5 & 419 & 6.2 & 12.88 & 4.35 \\
\hline
6 & 115 & 6.08 & 4.34 & 5.44 \\
\hline
7 & 21 & 4.76 & 4.76 & 6.86 \\
\hline
8 & 2 & 0 & 0 & 5.39 \\
\hline
\end{tabular}
\caption{p=0.6,q=0.96,$\rm{sup_{min}}$=0.3\%
\label{tab:06096}}
\end{center}
\end{table}

\begin{table}
\small
\begin{center}
\begin{tabular}{|c|c|c|c|c|}
\hline
Level & $|F|$ & $\sigma^+$ & $\sigma^-$ & $\rho$ \\
\hline
1 & 664 & 0.75 & 1.95 & 3.11 \\
\hline
2 & 1847 & 7.95 & 6.93 & 5.09 \\
\hline
3 & 1310 & 7.63 & 7.63 & 8.16 \\
\hline
4 & 864 & 8.91 & 16.31 & 14.29 \\
\hline
5 & 419 & 4.53 & 35.56 & 26.18 \\
\hline
6 & 115 & 0 & 53.91 & 46.81 \\
\hline
7 & 21 & 0 & 85.71 & 123.04 \\
\hline
8 & 2 & 0 & 100 & 0 \\
\hline
\end{tabular}
\caption{p=0.3,q=0.99,$\rm{sup_{min}}$=0.3\%
\label{tab:03099}}
\end{center}
\end{table}
Finally, since no single combination is the best in all respects, the
specific tradeoffs between privacy, accuracy and efficiency have to
be determined by the service provider to suit her requirements.  
Among the chosen candidates for $p,q$, moving to higher $p$ values gives
better accuracy, moving to lower $p$ values gives better privacy and 
moving to higher $q$ values gives better efficiency. 
But,the important point is that no matter which of these combinations is
chosen, they all provide 70-plus privacy, 80-plus accuracy, and slowdown
less than 5.

\subsection{Experiment Set 2: Real Dataset}
We conducted a similar set of experiments on the real dataset
(BMS-WebView-1~\cite{real}),  which had frequent itemsets of length
upto 4 for 0.3\% minimum support setting.  The results for this set of
experiments are shown in Table~\ref{tab:bms_1_results}. 
 
\begin{table}[ht]
\small
\begin{center}
\begin{tabular}{|c|c|c|c|c|c|c|c|c|c|c|}
\hline
p & q & BP & RP & $\Delta$ & $\sigma^+$ & $\sigma^-$ & $\rho$ & $Var(\sigma^+)$ & $Var(\sigma^-)$ & $Var(\rho)$ \\
\hline
0.8 & 0.96 & 92.7 & 71 & 4.8 & 3.21 & 3.67 & 3.21 & 2.34 & 1.67 & 0.67 \\
\hline
0.7 & 0.96 & 94.3 & 76 & 4.8 & 4.13 & 3.90 & 3.86 & 6.45 & 1.46 & 0.65 \\
\hline
0.6 & 0.96 & 95.8 & 81.4 & 4.8 & 5.05 & 5.28 & 4.80 & 11.96 & 4.47 & 0.42 \\
\hline
0.6 & 0.97 & 94.5 & 78 & 3.7 & 4.36 & 4.36 & 4.08 & 10.13 & 1.47 & 0.24 \\
\hline
0.7 & 0.97 & 92.7 & 73 & 3.6 & 2.98 & 3.67 & 3.34 & 3.05 & 1.67 & 0.36 \\
\hline
0.6 & 0.98 & 92.1 & 74 & 2.5 & 3.67 & 3.67 & 3.30 & 3.03 & 3.21 & 0.16 \\
\hline
0.5 & 0.98 & 94.3 & 79.5 & 2.4 & 4.36 & 4.82 & 4.35 & 3.50 & 11.53 & 0.20 \\
\hline
0.4 & 0.98 & 96.2 & 85.7 & 2.7 & 6.43 & 6.66 & 5.97 & 26.63 & 14.24 & 0.56 \\
\hline
0.4 & 0.99 & 93.2 & 79.3 & 1.6 & 5.28 & 5.05 & 4.78 & 15.40 & 16.50 & 1.34 \\
\hline
0.3 & 0.99 & 96 & 87 & 1.3 & 8.96 & 7.81 & 6.29 & 201.96 & 33.20 & 1.51 \\
\hline
\end{tabular}
\caption{Performance Results for Real Dataset
\label{tab:bms_1_results}}
\end{center}
\end{table}

The first point to note in Table \ref{tab:bms_1_results} is that the 
privacy and accuracy results are noticeably better than their counterparts
in the synthetic dataset experiment. On the other hand, the slowdown values
are marginally higher.  This is because Apriori is able to take full
advantage of the increased sparseness of this dataset, whereas 
EMASK still has to pay a price for the density increases that are an
outcome of the distortion process.  Secondly, there are a richer set of
tradeoffs that are available to the service provider to suit her
requirements. Finally, note again that setting very high values
of $q$, specifically $q=0.99$, while very attractive from privacy
and efficiency perspectives, results in high variance for the accuracy
measures. Here too, the errors are primarily incurred by the longer 
frequent itemsets.

\subsection{Summary}
Overall, our experiments indicate that by a careful choice of distortion
parameter settings, it is possible to simultaneously achieve satisfactory
privacy, accuracy, \emph{and efficiency}. In particular, they show that
there is a ``window of opportunity'' where these triple goals can be
all met.  The size and position of the window is primarily a function of
the database density and could be quite accurately characterized
with our estimation methods.

\comment{
Moving slightly around made a difference in desirability with respect
to one feature or other but the difference is not very high. But Moving
largely away from this window towards lower values of $p$, however,
results in skyrocketing errors, while increasing the value of $p$ will
result in significant loss of privacy.

We have conducted several other experiments with different market-basket
type databases and the results are consistent with those presented here.
}

\section{Related Work}
\label{sec:related}
The issue of maintaining privacy in association rule
mining has attracted considerable attention in the recent
past~\cite{very01,very01a,very01b,very01c,vaidya,clif02,gehrke02,lim03}.
However, to the best of our knowledge, none of these previous
papers have tackled the issue of efficiency in privacy-preserving mining.

The work closest to our approach is that of \cite{gehrke02,lim03}.  A set
of randomization operators for maintaining data privacy were presented
and analyzed in \cite{gehrke02}. New formulations of privacy breaches and
a methodology for limiting them were given in \cite{lim03}. The 
problem of large transactions resulting from distortion was also mentioned
in \cite{lim03}, but they addressed this problem from the perspective
of reducing storage and communication costs, but not runtime mining
efficiency. Specifically, they proposed a compression technique for
reducing the effective size of the distorted database.

Another difference is that EMASK (and MASK) support a notion of
``average privacy'', that is, they compute the probability of being
able to accurately reconstruct a \emph{random} entry in the database.
In contrast, \cite{gehrke02,lim03} evaluate the probability of
accurately reconstructing a \emph{specific} entry in the database.
In \cite{tech-report}, we present a detailed quantitative argument for
why it appears fundamentally unlikely that efficient mining algorithms 
can be designed to support this stronger measure of privacy.

Finally, the problem addressed in \cite{very01,very01a,very01b,very01c} is
how to prevent \emph{sensitive rules} from being inferred by the
data miner -- this work is complementary to ours since it addresses
concerns about \emph{output} privacy, whereas our focus is on the
privacy of the \emph{input} data.  Maintaining input data privacy is
considered in \cite{vaidya,clif02} in the context of databases that are
\emph{distributed} across a number of sites with each site only willing
to share data mining results, but not the source data.

\section{Conclusions}
\label{sec:conc}
In this paper, we have considered, for the first time, the issue of
providing efficiency in privacy-preserving mining. Our goal was to
investigate the possibility of simultaneously achieving high privacy,
accuracy and efficiency in the mining process. We first showed how the
distortion process required for ensuring privacy can have a marked
negative side-effect of hugely increasing mining runtimes. Then, we
presented our new EMASK algorithm that is specifically designed to
minimize this side-effect through the application of symbol-specific
distortion.  We derived simple but effective formulas for estimating
acceptable settings of the distortion parameters.  We also presented
a simple but powerful optimization by which all additional counting
incurred by privacy preserving mining is moved to the end of each pass
over the database.

Our experiments show that EMASK exploits a small window of
opportunity around the distortion combination ($p=0.4,q=0.98$) which
can simultaneously provide good privacy, accuracy and efficiency.
Specifically, less than 5 times slowdown with respect to Apriori in conjunction 
with 70-plus privacies and 80-plus accuracies, were achieved with these settings.  
In summary,
EMASK takes a significant step towards making privacy-preserving mining
of association rules a viable enterprise.

\comment{
\subsubsection*{Acknowledgements}
This work was supported in part by a Summer Research Fellowship from the
Centre for Theoretical Studies, Indian Institute of Science, and by
a Swarnajayanti Fellowship from the Dept. of Science \& Technology,
Govt. of India.
}

\jhcomment{Comment out all those references that are not being used.}

\bibliographystyle{plain}

\end{document}

%% file: cover.tex
\thispagestyle{empty}
\setcounter{page}{0}

\vspace*{1.5in}

\begin{center}

\begin{Large}
\begin{tabular} {l l}
Paper No:  &  {\bf 303} \\
& \\
Title: & On Addressing Efficiency Concerns \\
       & in Privacy-Preserving Mining\\
& \\
Authors: &  Shipra Agrawal, Vijay Krishnan, Jayant Haritsa \\
& \\
Contact Author: & Jayant Haritsa \\
& Supercomputer Education and Research Centre \\
& Indian Institute of Science \\
& Bangalore~560012, INDIA  \\
& \\
& Email: haritsa@dsl.serc.iisc.ernet.in \\
& Tel:  (+91)-80-2932793 \\
& Fax:  (+91)-80-3602648 \\
& \\
\end{tabular}
\end{Large}
\end{center}